# The Dengue risk transmission during the FIFA 2014 World Cup

P. S. Lucio[+1,2], N. Degallier[1], M. H. C. Spyrides[2], Cláudio M. S. e Silva[2], J. C. B. da Silva[2],

H. J. F. da Silva[2], G. Máximo[2], W. Junior[2], M. Mesquita[3]

[1] Laboratoire d'Océanographie et du Climat, Expérimentations et Approches Numériques (LOCEAN-IPSL), IRD UMR 182, Paris France : nicolas.degallier@ird.fr

[2] Universidade Federal do Rio Grande do Norte (UFRN), Centro de Ciências Exatas e da Terra (CCET), Campus Universitário - Lagoa Nova, 59078-970 - Natal – RN, Brazil : pslucio@ccet.ufrn.br  psllod@locean-ipsl.upmc.fr

[3] Uni Research, Bjerknes Centre for Climate Research – Bergen - Norway

[+]Visiting Scientist

Corresponding author: P. S. Lucio, Universidade Federal do Rio Grande do Norte (UFRN), Centro de Ciências Exatas e da Terra (CCET), Campus Universitário - Lagoa Nova, 59078-970 - Natal – RN, Brazil; , e-mail : pslucio@ccet.ufrn.br

**ABSTRACT:** Dengue is a persistent threat to Brazilians. The twentieth FIFA World Cup will take place in Brazil in June and July this year. This paper warning that Dengue fever could be not a significant problem in competition locations. Temperature is a very strong predictor of suitability for Dengue and it seems evident that relative humidity measurements are less satisfactory in biological (ecological, entomological or even in epidemiological studies) work as indicator of atmospheric factors influencing evaporation than are measurements of vapor pressure deficit. In fact, the vapour pressure deficit is a much more sensitive indicator of water vapor conditions of the atmosphere and undergoes greater variations for temperature changes than does the relative humidity. Further, relative humidity alone gives no indication of the rate of evaporation while the vapour pressure deficit alone does give an indication of the evaporation rates, then areas having the same vapour pressure deficit do influence evaporation rates in the same way whether temperature are identical or not. It is the reason why vapour pressure deficit was chosen in this study. Human and mosquito populations associated with precipitation are good predictors of Dengue in suitable places but don't capture the whole picture. The determinants of Dengue virus transmission is composition of suitable climate, susceptible people and competent mosquitos virus interactions. For the 2014 World Cup twelve host cities the Dengue risk transmission will be not close to its peak when matches will be played, because this is a non epidemic period overall the brazilian territory. Despite the low risk – it always exists - the Brazilian authorities must communicate this fact and what protective measures FIFA World Cup fans should be take. In fact, the risk of an outbreak of dengue fever during the upcoming soccer World Cup in Brazil is not serious enough to warrant a high alert in the host cities, according to an early warning system for the disease.

***Key-words:*** *Aedes aegypti, epidemiology, climatic factors, mosquito development rate, vectorial capacity.*

**Significance Statements**

Dengue is a viral infection that can produce a severe fever and symptoms that may require hospitalization. It is transmitted between humans by the urban-adapted, day-biting *Aedes* mosquitoes and is therefore a particular problem in towns and cities. To explore this risk, one has assessed the potential levels of exposure by a climate-driven model for dengue risk transmission in Brazil and records of its seasonal variation at the key sites. Like the weather, it is unworkable to forecast the precise situation with regard to dengue in Brazil in 2014. One can, however, make informed guesses on the basis of averaged records of dengue in previous years. For the areas around the World Cup stadiums, these records show that the main dengue season will have passed before the World Cup is held in June and July. Unfortunately, the risk remains, even this is low but not negliegible, during these months in the Brazilian north and northeast.

**Introduction**

The tropical seasonal climate of Brazil provides the optimum conditions for the *Aedes aegypti* mosquito to breed. The majority of cases occurr in the rainy and autumn seasons between January and May. The general upwards trend in dengue fever infection rates demonstrates the difficulties of controlling the dengue vector with traditional chemical control methods implemented in Brazil.

The purpose of this paper is to discuss the Simon Hay (2013) paper, exploring the relationship between weather variations and the *Aedes aegypti*-based Dengue risk transmission. So, a mathematical model (Lucio et al, 2014), which simulates the risk of Dengue fever transmission under a given climate – a climate-dependent model based on the development phases of the Dengue vector – was applied for the 2014 World Cup twelve host cities. In Lucio et al (2014) the effect of temperature and relative humidity on the ability of *Ae. aegypti* to transmit Dengue virus is assessed as a possible explanation for the seasonal variation in the incidence of the virus in some endemic and epidemic brazilian cities with contrasting climate (Barcellos and Lowe, 2013). In fact, based on the favorability in terms of the vectorial capacity, the Dengue risk transmission will be not close to its peak when matches will be played in none of the host cities, because this is a non epidemic period (Massad et al., 2008) overall the brazilian territory.

**Dengue and Dengue Vector**

The *Ae. aegypti* mosquito is the primary vector of Dengue. The virus is transmitted to humans through the bites of infected female mosquitoes. After virus incubation for 4–10 days, an infected mosquito is capable of transmitting the virus for the rest of its life. Infected humans are the main carriers and multipliers of the virus, serving as a source of the virus for uninfected mosquitoes. Patients who are already infected with the Dengue virus can transmit the infection (for 4–5 days; maximum 12) via *Aedes* mosquitoes after their first symptoms appear. The *Ae. aegypti* lives in urban habitats and breeds mostly in man-made containers. Unlike many other mosquitoes *Ae. aegypti* is a daytime feeder; its peak biting periods are early in the morning and in the evening before dusk. The *Ae. aegypti* female may bite multiple people during each feeding period. The Dengue fever is severe desease and the symptoms usually last for 2–7 days, after an incubation period of 4–10 days, after the bite from an infected mosquito.

**Dengue Vector and Weather**

Increases in the temperature of a system results from increases in the kinetic energy of the system (Focks et al., 1993a, 1993b). This has several effects on the mosquito development rates. Each day the rate of development and survivorship (or mortality) for each cohort is calculated. Survivorship and development requirements are life stage-dependent. The metamorphosis is dependent on daily cumulative development and other life stage-specific factors. Many entomological models use day-degree or temperature average models that assume development rate is proportional to temperature. This suits modeling development within a limited temperature range. Sharpe and DeMichele (1977) developed an enzyme kinetics model appropriate for use over a significantly wider temperature range. In the model, a single rate-controllingenzyme determines development rates. Development accumulates each day and is complete when it reaches a specified level. According to Focks et al. (2000) an important factor in *Ae. aegypti* Dengue transmission risk modeling is the influence of temperature on the extrinsic incubation period (EIP) of the virus in the female. At higher temperatures, infected females become infectious through viral dissemination at a significantly faster rate. Thus, the probability of an infected female living long enough to become infectious goes up significantly with temperature. Specifically, the probability of surviving the incubation period using values for EIP from Focks et al. (1997), given that females incubating virus at the higher temperature are 2.64 times more likely to survive long enough to potentially infect human hosts. Besides, *Ae. aegypti* survival (or mortality) rates are temperature dependent. These rates are based on laboratory studies and anecdotal field observations, which could vary from what occurs in the field. Moisture also affects mosquito survival. Larvae require water for survival, but pupae do not, and vapor pressure deficit affects adult mortality (Christophers, 1960, Focks and Chadee, 1997).

**Description of the Mathematical Model**

In the model one considers that development accumulates each day and is completed when the cumulative development is greater than $cdi = 0.95$ (the cumulative development factor for first life cycle) for immature stages. One also considers that *Ae. aegypti* survival rates are temperature-dependent. Daily minimum and maximum temperatures, linearly approximated from the monthly values, determine survivorship rates. These rates are based on laboratory studies and subjective field observations, which could vary from what occurs in the field. At temperature extremes one set the survival factor to 0.00 (or 0.05), since these studies indicate that mosquitoes at any life stage are unlikely to survive extreme temperatures. Moisture also affects mosquito survival. Studies show eggs can survive desiccation for many months, but for hatching, water is necessary. We have simplified our model and assume that the mature egg will hatch if water is present in the ovipositional container. Larvae require water for survival, but pupae do not, and vapor pressure deficit affects adult mortality (Christophers, 1960). The one day survival-base probability is considered constant. The model calculate the development rate at a given temperature $T(^{o}K)$ per hour based on the development rate at the ideal mean temperature on a day for larval and pupal development, the value of the enthalpy for the activation of the enzyme catalytic reaction (in cal.mol$^{-1}$) and the perfect gas constant. So, each monthly cohort development period ($\tau$ in days) is given as the following equations, where the model parameters were stablished with the procedure adopted by Wagner *et al.* (1984). In **Table 1** and **Table 2** are given detailed descriptions of the mosquito life parameters and their dependence on climate.

The basic idea of the dengue risk transmission modeling takes into account the fact that it is easier to achieve a level of larvae ot pupae per inhabitant given a warm and humid climate than a cold and dry one. Initially we defined a transmission risk index by dividing the number of pupae per host by the maximum reproduction rate of the mosquitoes. The issue is that the index thus obtained is not so easy to understand. Thus, one builds an alternative risk index closer to what one wants to quantify, *i.e.* the environment required (number of pupae/human host) in a given climate, to be favorable to an epidemic transmission of dengue.

This risk index is based on the suitability of the climate for completion of the mosquito cycle and for transmission of the virus from an infected host to a susceptible one. Hence, to design the model one starts from the relationship between the dengue basic reproduction rate ($R_0$) and the number of pupae per host in the environment by the classical model (*cf*. Degallier *et al*., 2010)

**Supporting Physical Reasoning**

Vapor pressure is a measurement of the amount of moisture in the air. It is technically the pressure of water vapor above a surface of water. The saturation vapor pressure (SVP) is the pressure of a vapor when it is in equilibrium with the liquid phase. It is solely dependent on the temperature. As temperature rises the saturation vapor pressure rises as well. The vapour pressure deficit (VPD) is the difference (deficit) between the amount of moisture in the air and how much moisture the air can hold when it is saturated. The VPD calculation is an improvement over relative humidity measurement alone because VPD takes into account the effect of temperature on the water holding capacity of the air. "The constraint under which an organism is placed in maintaining a water balance during temperature changes is much more clearly shown by noting the VPD than by recording the relative humidity." (Souce: Anderson, D. B. 1936).

The VPD is the measure of the drying power of the atmospheric air. It is calculated as a function of temperature and relative humidity. Moreover, Anderson (1936) explained that giving a rise of temperature from 20$^{o}$C to 30$^{o}$C, assuming no change in the vapour pressure of the atmosphere, brings a change of less than 32% in the relative humidity, but a variation of more than 370% in the VPD. From this issue, it seems evident that relative humidity measurements are less satisfactory in biological (ecological, entomological or even in epidemiological studies) work as indicator of atmospheric factors influencing evaporation than are measurements of VPD. In fact, it is a much more sensitive indicator of water vapor conditions of the atmosphere and undergoes greater variations for temperature changes than does the relative humidity. Further, relative humidity alone gives no indication of the rate of evaporation while the VPD alone does give an indication of the evaporation rates, then areas having the same VPD do influence evaporation rates in the same way whether temperature are identical or not (Focks et al., 2006).

**Material**

Source of the Meteorological Dataset: Brazilian Meteorological Institute (Instituto Nacional de Meteorologia; INMET in BDMEP, http://www.inmet.gov.br/portal/). Source of the Dengue Dataset: Brazilian public health surveillance system (Sistema de Informação de Agravos de Notificação; SINAN, http://dtr2004.saude.gov.br/sinanweb/).

**Concluding Remarks**

As showed in **Figure 1** the Dengue risk transmission will be not close to its peak when matches will be played in the host cities. This is corroborated by **Figure 2**. The 2014 risk forecast (**Table 3**) was obtained via the climate forecast of temperature and relative humidity was obtained via the Holt-Winters' double exponential smoothing technique, using a 30 years dataset collected in the host cities.

This study contributes to a better understanding of the dynamics of Dengue based on the mosquitos' subsistence under tropical climate conditions. The Lucio et al. (2014) generic model can be carried out for arbovirus diseases, coupling geographic models, which describe the vector distribution based on environmental local conditions, and considering that climate is an important geographic determinant of vectors activity. For the 2014 World Cup concerned areas these records show that the main Dengue season will have passed before June and July. Unfortunately, as a tropical country the risk is never zero during all months in the year. As mentioned by Simon Hay (2013): "The World Cup is an opportunity to evaluate the uptake of these new public-health information systems and their utility both to individuals and the authorities. Crucially, if they can provide timely feedback on the effectiveness of preventive measures for the authorities on the ground, that could prompt yet further responses." As a final comment, the 2014 World Cup will be played out in stadiums in the twelve host cities. All the stadiums in these twelve cities are therefore located in areas of low Dengue risk transmission. Climate or weather predictions are always along with a percentage of uncertainties and the same is true for the Dengue risk transmission.

Clinical and epidemiological aspects of Dengue in Brazil have stimulated the interest of researchers as well as national and international public health agencies, in view of the importance of identifying factors that determine the differences on individual and collective forms expressing these infections with the aim to improving its treatment and control.

The presence of Dengue and its vector in a given region depends on climatic and environmental characteristics to which this region is subjected. The regions located in the intertropical zone, like Brazil, are favorable to the virus transmission throughout the year. However, in the cities hosting the World Cup, June and July are the months that have lower risk indicators. This can be seen in **Figure 1** which shows the monthly behavior of the risk of Dengue in 12 capital cities that will host the event.

It is observed that the risk of Dengue is lower during the months in the cities where lower DPVS values are recorded. This is due to the fact that in the months of June and July the air temperature decreases thus leading to low levels of vapor pressure, unfavorable for mosquito survival.

Of the twelve cities which will host the World Cup, eleven have reduced risk compared to their historical prevalence, in June and July months. Only Manaus, capital of Amazonas, located in the Northern region of the country is at average risk considering the region historical registers with an elevation trend from June, but the most critical periods are August, September and October, coinciding with the rise of SVPD during these months.

The risk prediction for June and July 2014, the period of the event, for the 12 cities, are shown in Table 1 in descending order considering each of the regions of Brazil. It appears that the most critical city is Manaus risk prediction for June and July, about 80%, followed by Fortaleza (73%) and Natal (70%). The lowest risks were predicted for the more Southern cities: Porto Alegre and Curitiba, with approximately 26%.

In summary, it was found that the increased risk of Dengue, most often, is associated with an increased SVPD, what happens in the period from the last to the first months of the year due to the predominance of the rainy season, which is a factor conducive to increased proliferation of mosquitoes in the studied cities.

While recognizing that at the moment of the World Cup most host cities has a history of lower Dengue vulnerability, actions should be taken to mitigate the risk of disease. In this sense, recommendations related to preventing the accumulation of water must be strengthened and monitored, such as attention to using dish in potted plants, animal drinkers and open water deposits, cleaning the gutters and ledges at weekly intervals, maintain storage containers closed, guarding empty bottles upside down and deliver unused tires for cleanup crews. Dengue is a problem that worsens with the socio-cultural conditions of a society, therefore, educational policies are of utmost importance for monitoring and controlling the disease.

**Drawing Relevant Public Health Conclusions**

The aim of this manuscript is to minimise the number of locally acquired cases of dengue in Brazil by strengthening and sustaining risk based surveillance, prevention and control measures for both human cases and the mosquitoes that carry the dengue virus. So, the brazilian authorities must aim to achieve this by improving disease surveillance, enhancing and coordinating mosquito surveillance, prevention and control measures and by educating the community, industry and relevant professional groups. They have to ensure the timely detection and reporting of all suspected dengue cases, support effective and timely control methodologies to prevent local transmission of dengue, establish a state-wide surveillance program for the detection of dengue vectors, reducing the spread of the dengue vectors across the main cities.

| Parameter | Notation | Climate Dependence |
|---|---|---|
| Duration of the larval stage | $\tau_l$ (day) | $\tau_l = \dfrac{1}{24(rol/cdi)} \dfrac{T_{ref}}{tempK} f_{larval}$ with $f_{larval} = e^{(dhal/R.(1./tempK-1./T_{ref}))}(1+e^{(dhhl/R.(1./Tl50ht-1./tempK))})$ |
| Duration of the pupal stage | $\tau_p$ (day) | $\tau_p = \dfrac{1}{24(rop/cdi)} \dfrac{T_{ref}}{tempK} f_{pupal}$ with $f_{pupal} = e^{(dhap/R*(1./tempK-1./T_{ref}))}$ |
| Duration of the immature stage | $\tau_I$ (day) | $\tau_I = \tau_l + \tau_p$ |
| Basic immature (larval and pupal) mortality rate | $\mu_I^0$ (day$^{-1}$) | $\mu_I^0 = \max\left\{.01+.99e^{\left(-\frac{(T-5)}{2.7035}\right)}, .01+.99e^{\left(\frac{-(42-T)}{3.}\right)}\right\}$ |
| Adult mortality rate or adult mosquitoes' daily mortality rate | $\mu_A$ (day$^{-1}$) | $\mu_A = -\ln[0.89 f(T) g(\delta_e)]$ with $f(T) = \min\left\{1, \max\left(0; \dfrac{T-5}{5}; \dfrac{43-T}{2}\right)\right\}$ and $g(\delta_e) = \min\left\{1; \max\left(0.6; 1-0.4\dfrac{\delta_e - 10}{20}\right)\right\}$ |
| Duration of the time between emergence and the first bite | $\tau_1$ (day) | $\tau_1 = \dfrac{1}{24roj} \dfrac{T_{ref}}{tempK} f_{1st}$ with $f_{1st} = e^{(dhag/R*(1./tempK-1./T_{ref}))}$ |
| Duration of the gonotrophic cycle | $\tau_g$ (day) | $\tau_g = 0.58\tau_1$ |
| Duration of the extrinsic (vector) incubation period | $\tau_{EIP}^0$ (day) | $\tau_{EIP}^0 = \dfrac{1}{24(roip0/cdi)} \dfrac{T_{ref}}{tempK} f_{EIP}$ with $f_{EIP} = e^{(dhaeip0/R*(1./tempK-1./T_{ref}))}$ |
| Delay between the bite infecting the vector and its first infective bite | $\tau_{EIP}$ (day) | $\tau_{EIP} = \left[\left(\dfrac{t_{EIP}^0}{t_g}\right) + 1\right] t_g$ |

**Table 1a:** The parameters that are climate dependent of the mosquito life cycle parameters in temperature $T(^oC)$ and saturated vapour pressure deficit $\delta_e (mb)$.

| Parameter | Notation | Range in the control model |
|---|---|---|
| Survival at emergence | $p_e$ | $p_e = 0.91$ or $p_e = 0.87$ or $p_e = 0.83$ or ... |
| Proportion of females at emergence | $\rho$ | $\rho = 0.5$ |
| Mean number of different hosts bitten for a complete blood meal - the mosquitoes' daily biting rate | $\mu_{host}$ | $\mu_{host} = 1.8$ or $\mu_{host} = 2.0$ or ... |
| Mean number of bites per day per vector and per host | $\alpha$ (day$^{-1}$) | 0–0.5 |
| Proportion of bites of infectious vectors on susceptible hosts leading to host infection | $b$ | 0–1 |
| Proportion of bites of susceptible vectors on viremic host leading to vector infection | $c$ | 0–1 |
| Number of vector by hosts |  | 0–3 |
| Inverse of the duration of host viraemia | $1/c$ (day$^{-1}$) | 0–0.5 |
| intrinsic (host) incubation period | $t_v$ (day) | 1–11 |
| Duration of viraemia in hosts - intrinsic (host) incubation period | $t_v$ (day) | $t_v = 6$ |

**Table 1b:** The parameters that are non-critical: as they only act proportionally on the climate suitability, a change in their value does not change the pattern of variation of suitability.

| Parameter description | Notation | Range in the model |
|---|---|---|
| Mean number of bites per day per vector and per host | $a$ | 0–0.5 day$^{-1}$ |
| Proportion of bites of infectious vectors on susceptible hosts leading to host infection | $b$ | 0–1 |
| Proportion of bites of susceptible vectors on viremic host leading to vector infection | $\gamma_h$ | 0–1 |
| Number of vector by hosts | $m$ | 0–3 |
| Initial epidemic growth rate | $r$ | 0.33 |
| Inverse of the duration of host viraemia | $\gamma_h^{-1}$ | 0–0.5 day$^{-1}$ |
| Mean host infectious period | | 3-7 days |
| Mortality rate of the vectors | $\mu_v$ | 0–0.2 day$^{-1}$ |
| Mean adult mosquito lifespan | $\mu_v^{-1}$ | 6-15 days |
| Mean extrinsic (vector) incubation period | $\tau_e$ | 1–16 days |
| Mean intrinsic (host) incubation period | $\tau_i$ | 1–11 days [(4-7) days] |

**Table 2:** The parameters of the deterministic aggregated dynamical model used in the estimation of the reproduction number.

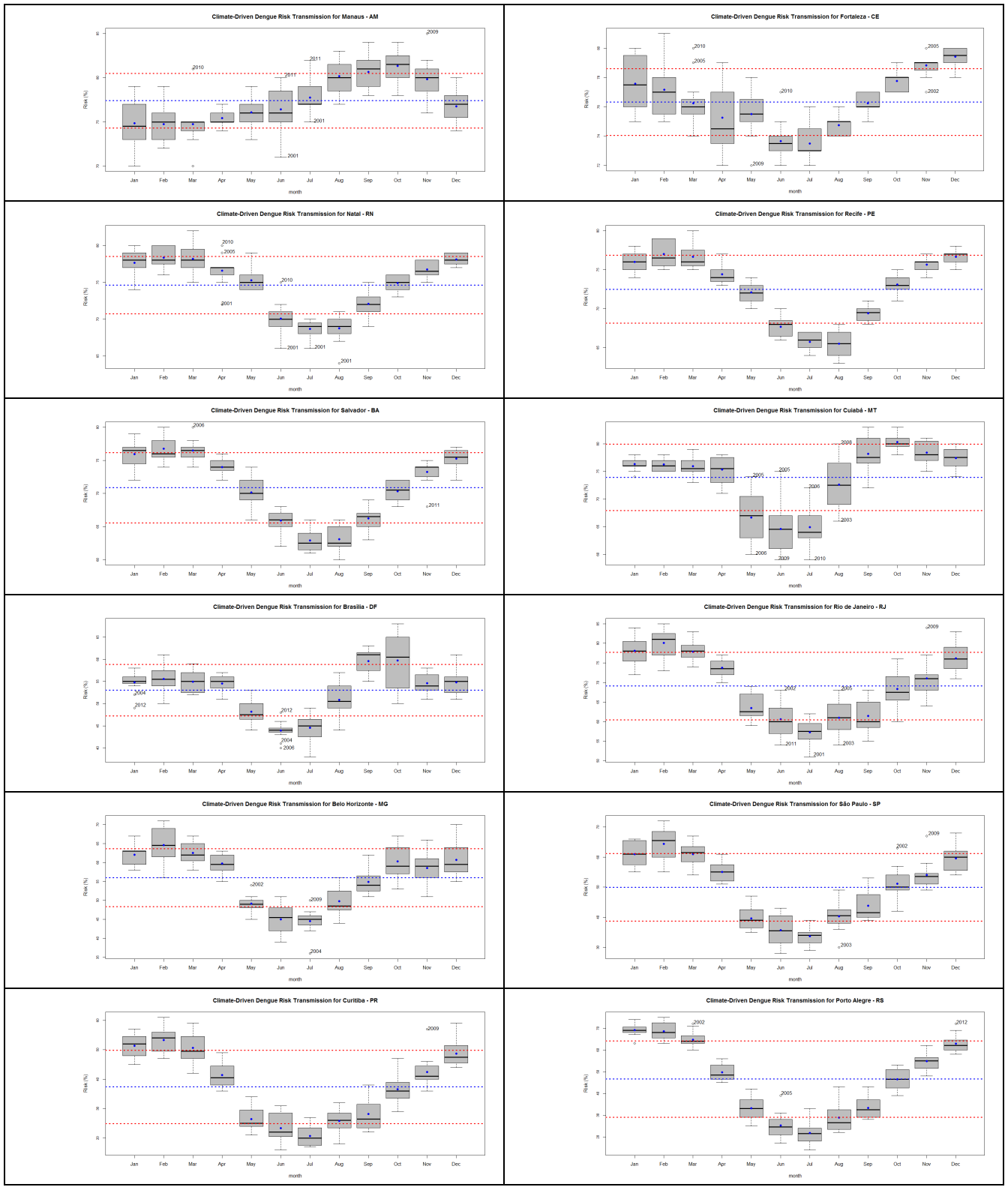

**Fig. 1:** The 12-years monthly Box-Plot for the Seasonal Dengue Maximum Risk Transmition (based on the *Ae. aegypti*). In dashed blue the annual average, the dashed red lines are the one standard deviation interval around the annual average and the blue dots are the monthly mean. Both variables with the identification of the very anomalous outlined years.

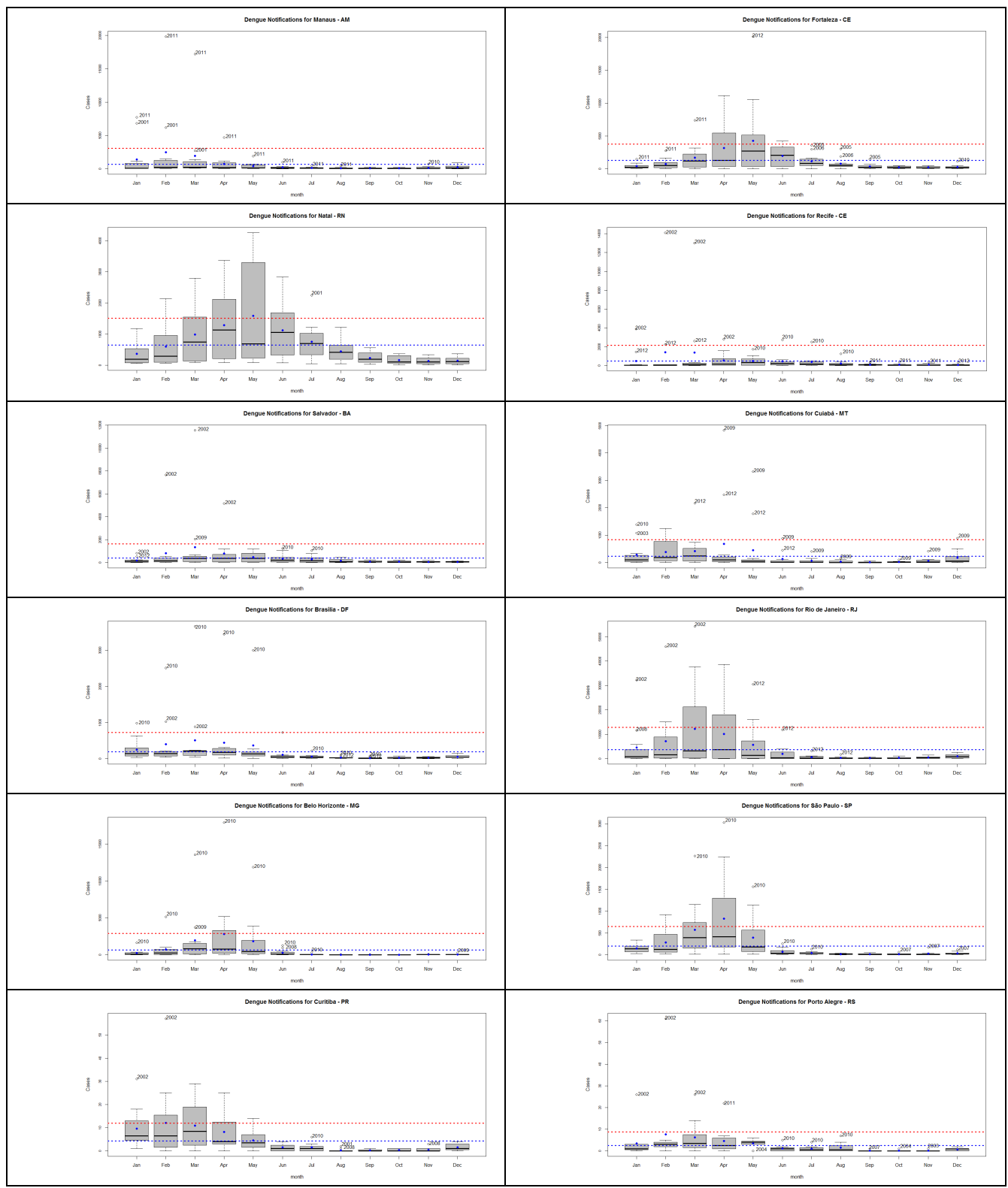

**Fig. 2:** The 12-years monthly Box-Plot for the DEN notifications by sick patient residence, analysed at a municipality level - reported Dengue cases by month (2001-2012). The seasonal distribution of Dengue in an average year in the locations of the 2014 World Cup stadiums. In dashed blue the annual average, the dashed red lines are the one standard deviation interval

| Brazilian Region | Capital City - State | jun/14 | | | jul/14 | | |
|---|---|---|---|---|---|---|---|
| | | lower | forecast | upper | lower | forecast | upper |
| **North** | Manaus - AM | 0,73 | 0,80 | 0,86 | 0,73 | 0,80 | 0,87 |
| **Northeast** | Fortaleza - CE | 0,66 | 0,73 | 0,79 | 0,66 | 0,72 | 0,79 |
| | Natal - RN | 0,62 | 0,70 | 0,78 | 0,60 | 0,68 | 0,76 |
| | Recife - PE | 0,64 | 0,68 | 0,72 | 0,61 | 0,66 | 0,70 |
| | Salvador - BA | 0,60 | 0,65 | 0,70 | 0,57 | 0,62 | 0,67 |
| **Midwest** | Cuiabá - MT | 0,58 | 0,66 | 0,73 | 0,58 | 0,66 | 0,73 |
| | Brasília - DF | 0,36 | 0,46 | 0,57 | 0,36 | 0,47 | 0,57 |
| **Southeast** | Rio de Janeiro - RJ | 0,48 | 0,62 | 0,76 | 0,47 | 0,62 | 0,76 |
| | Belo Horizonte - MG | 0,37 | 0,47 | 0,57 | 0,38 | 0,48 | 0,58 |
| | São Paulo - SP | 0,25 | 0,38 | 0,52 | 0,25 | 0,39 | 0,53 |
| **South** | Porto Alegre - RS | 0,15 | 0,27 | 0,40 | 0,12 | 0,25 | 0,37 |
| | Curitiba - PR | 0,11 | 0,26 | 0,40 | 0,12 | 0,26 | 0,41 |

**Tab. 1:** The 2014 forecast for the Dengue Maximum Risk Transmition (based on the *Ae. aegypti*) with their respective 95% confidence intervals.